\documentclass[a4paper,11pt]{article}

\usepackage{jinstpub} 

\RequirePackage{comment}
\RequirePackage{physics}
\title{\boldmath The calibration house in JUNO}

\author[a,1]{J. Hui,\note{Corresponding author.}}
\author[b]{R. Li,}
\author[b]{Y. Wu,}
\author[a,2]{T. Zhang,\note{Corresponding author.}}
\author[b]{Z. Chen,}
\author[b]{A. Freegard,}
\author[b]{J. Huang,}
\author[b]{H. Lai,}
\author[b]{Y. Liao,}
\author[a,b,c]{J. Liu,}
\author[b]{Y. Meng,}
\author[a]{A. Takenaka,}
\author[b]{Z. Xiang,}
\author[b]{P. Zhang,}
\author[a]{Y. Zhang}

\affiliation[a]{Tsung-Dao Lee Institute, Shanghai Jiao Tong University, \\Shanghai 201210, China}
\affiliation[b]{School of Physics and Astronomy, Shanghai Jiao Tong University, \\MOE Key Laboratory for Particle Astrophysics and Cosmology, \\Shanghai Key Laboratory for Particle Physics and Cosmology,
\\Shanghai 200240, China}
\affiliation[c]{New Cornerstone Science Laboratory, Tsung-Dao Lee Institute, Shanghai Jiao Tong University, \\Shanghai 201210, China}

\emailAdd{huijiaqi@sjtu.edu.cn}

\abstract{
As an auxiliary system within the calibration system of the Jiangmen Underground Neutrino Observatory, a calibration house is designed to provide interfaces for connecting the central detector and accommodating various calibration sub-systems. Onsite installation has demonstrated that the calibration house interfaces are capable of effectively connecting to the central detector and supporting the installation of complex and sophisticated calibration sub-systems. Additionally, controlling the levels of radon and oxygen within the calibration house is critical. Radon can increase the experimental background, while oxygen can degrade the quality of the liquid scintillator. The oxygen concentration can be maintained at levels below 10 parts per million, and the radon concentration can be kept below 15 mBq/m$^3$. This paper will provide detailed information on the calibration house and its methods for radon and oxygen concentration control.
}


\arxivnumber{to be filled} 

\begin{document}
\maketitle

\flushbottom

\section{Introduction}
\label{sec:intro}
The Jiangmen Underground Neutrino Observatory (JUNO) is a Liquid Scintillator (LS) detector located in Kaiping City, Jiangmen City, Guangdong Province, China, 53~km away from both the Yangjiang and Taishan nuclear power plants. It is designed to determine the neutrino mass ordering by precisely measuring the energy spectrum of reactor neutrinos. The detector will also detect atmospheric neutrinos, solar neutrinos, geoneutrinos, and supernova neutrinos~\cite{JUNO:2021vlw}. Data taking is expected to begin in 2025. The JUNO Central Detector (CD) consists of an acrylic spherical vessel with an inner diameter of 35.4~m, filled with 20~kton of LS as the target material, designed to detect Inverse Beta Decay (IBD). Approximately 17,600 20-inch and 25,000 3-inch Photomultiplier Tubes (PMTs) are mounted on a Stainless Steel (SS) structure outside the acrylic vessel, tasked with detecting photons produced by IBD events occurring inside the CD~\cite{Abusleme:2024aa}. To achieve the physics goals, an effective energy resolution better than 3\% and an energy scale uncertainty below 1\% are required~\cite{Zhan:2009rs}.

\begin{figure}[!htbp]
    \centering
    \begin{minipage}{0.45\textwidth}
        \centering
        \includegraphics[width=\textwidth]{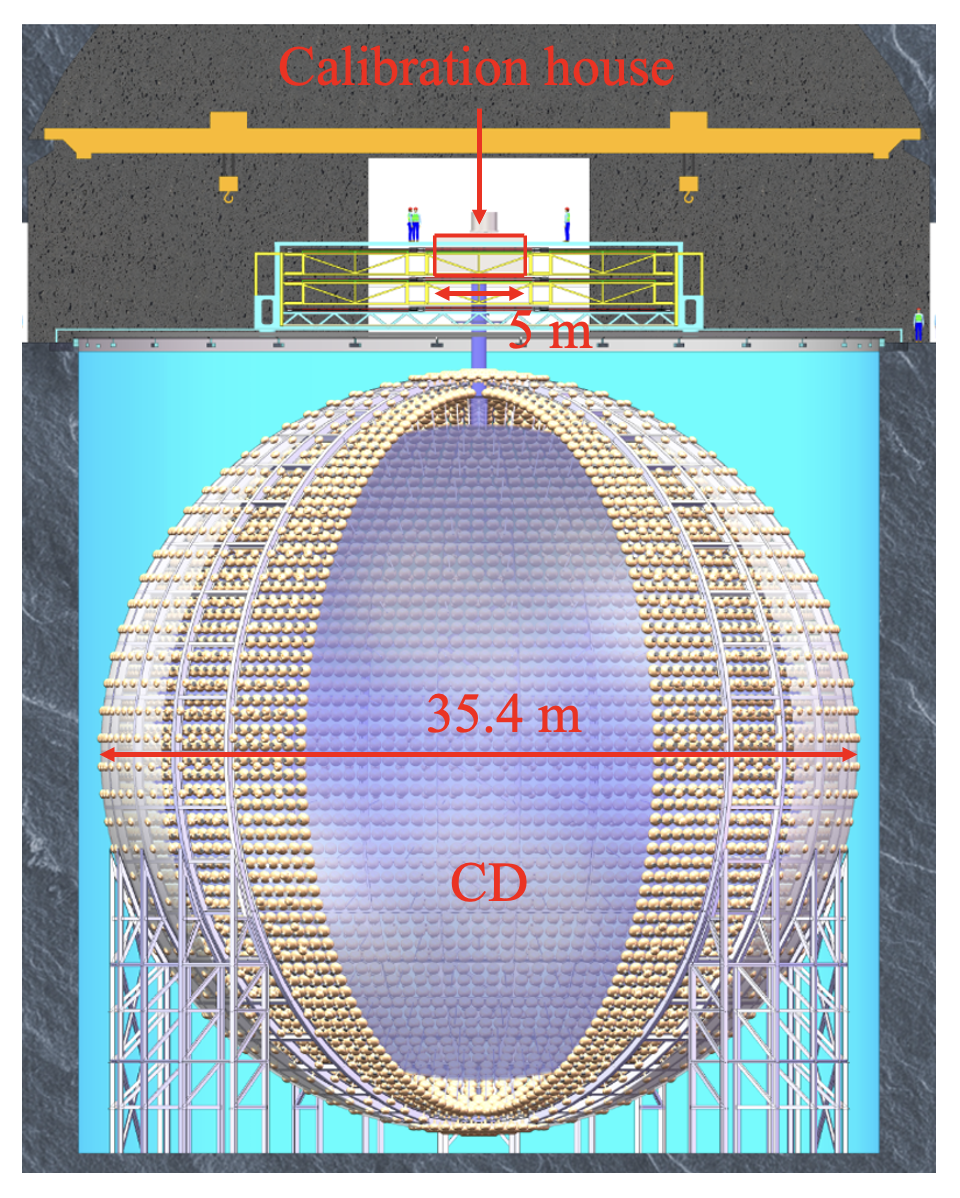}
        \label{fig:left}
    \end{minipage}%
    \hspace{0.05\textwidth}
    \begin{minipage}{0.45\textwidth}
        \centering
        \includegraphics[width=\textwidth]{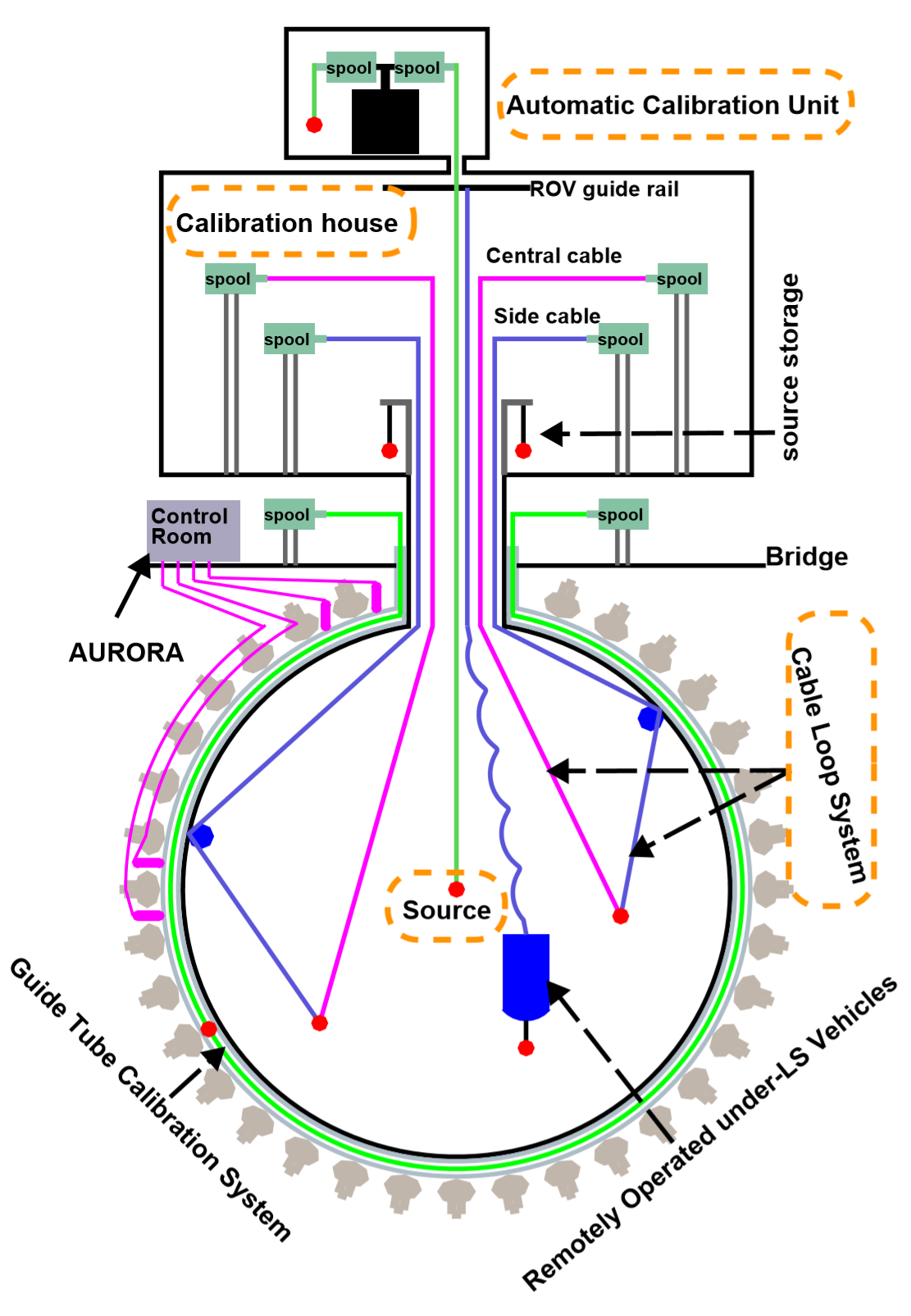}
        \label{fig:right}
    \end{minipage}
    \caption{The left plot is an engineering diagram of the CD and calibration house. The right plot is a schematic of the calibration system (not to scale)~\cite{JUNO:2020xtj}. It consists of the calibration source, ACU, the CLS, the GTCS, the ROV, and auxiliary system.}
    \label{JUNO_calibration_system}
\end{figure}

To meet the requirements of correcting both energy non-linearity and position-dependent non-uniformity of the CD, the calibration system must be capable of deploying multiple radioactive sources and laser sources to designated locations~\cite{JUNO:2020xtj,Zhang:2018yso,Takenaka:2024ctk}. Consequently, four independent subsystems, as shown in Figure~\ref{JUNO_calibration_system}, each with different coverage areas and usage frequencies, have been designed: the Automatic Calibration Unit (ACU)~\cite{Hui:2021dnh}, the Guide Tube Calibration System (GTCS)~\cite{Guo:2019fkf,Guo:2021ugw}, the Cable Loop System (CLS)~\cite{Zhang:2020grf}, and the Remotely Operated under-LS Vehicle (ROV)~\cite{Feng:2018xad}.

The ACU is a fully automated mechanical system capable of weekly deployments of various radioactive sources and a laser source along the CD’s central axis. The GTCS enables the monthly deployment of a radioactive source along a given longitude on the CD’s outer surface. The CLS is designed for monthly placement of multiple radioactive sources at designated positions in a given vertical plane inside the CD. The ROV allows annual deployment of radioactive sources to nearly any location inside the CD. Additionally, several auxiliary systems have been developed: the Ultrasonic System (USS)~\cite{Zhu:2019vay,Teng:2022usb,Zhu:2024keu}, the Charge-Coupled Device (CCD) camera positioning system, the automated source changing mechanic, and the calibration house. The USS and the CCD camera positioning system both serve as positioning tools for radioactive sources deployed by the CLS. The automated source changing mechanic can automatically change various radioactive sources for the CLS. The calibration house provides the interface to the central detector and serves as an installation chamber for the calibration subsystems.

This paper is organized as follows. In Section~\ref{sec:requirements}, the design requirements of the calibration house are introduced. Section~\ref{sec:design} discusses the design of the calibration house. Tests of the calibration house are presented in Section~\ref{sec:test}. The final section summarizes the paper.

\section{Design requirements of the calibration house}
\label{sec:requirements}
The calibration house is designed to provide interfaces for connecting the CD and accommodating various calibration sub-systems. Due to its direct connection with the CD, the concentrations of radon (which contributes to detector background) and oxygen (which affects the LS performance) in the calibration house must also be strictly controlled. Therefore, the basic design requirements for the calibration house are as follows: 

\begin{itemize}
\item The calibration house must have sufficient interior space to accommodate the CLS and its automated source changing mechanic.
\item Both external interfaces (flanges for electric feedthroughs, flush pipes, the ACU, the ROV, the CD, and liquid level sensor during filling stage) and internal interfaces (fixtures for the CLS and monitoring camera installations) are required.
\item The overall leak rate of the calibration house must be less than 4~$\times$~10$^{-5}$~mbar$\cdot$L/s at 1500~Pa gauge pressure~\cite{leak_rate}.
\item The oxygen and radon concentration inside the calibration house must be maintained below 10 Parts Per Million (ppm) and 20~mBq/m$^3$, respectively~\cite{JUNO:2021kxb}.
\item A glove box is required for transferring calibration sources and auxiliary devices, as well as for handling exceptions, while preserving the overall air-tightness of the calibration house.
\item To minimize radioactive background and facilitate maintenance, the CLS electronics inside the calibration house must be connected to a control box located outside the calibration house via electric cables. Furthermore, the CLS electronics must be insulated from the calibration house.
\end{itemize}

\section{Design of the calibration house}
\label{sec:design}
In this section, the design of the interface and glove box, the distribution and grounding of electric cables, and the control strategies for oxygen and radon levels inside the calibration house will be introduced sequentially.

\subsection{Design of the interface}
\label{sec:design_interface}

\begin{figure}[!hbtp]
  \centering
  \includegraphics[width=0.8\textwidth]{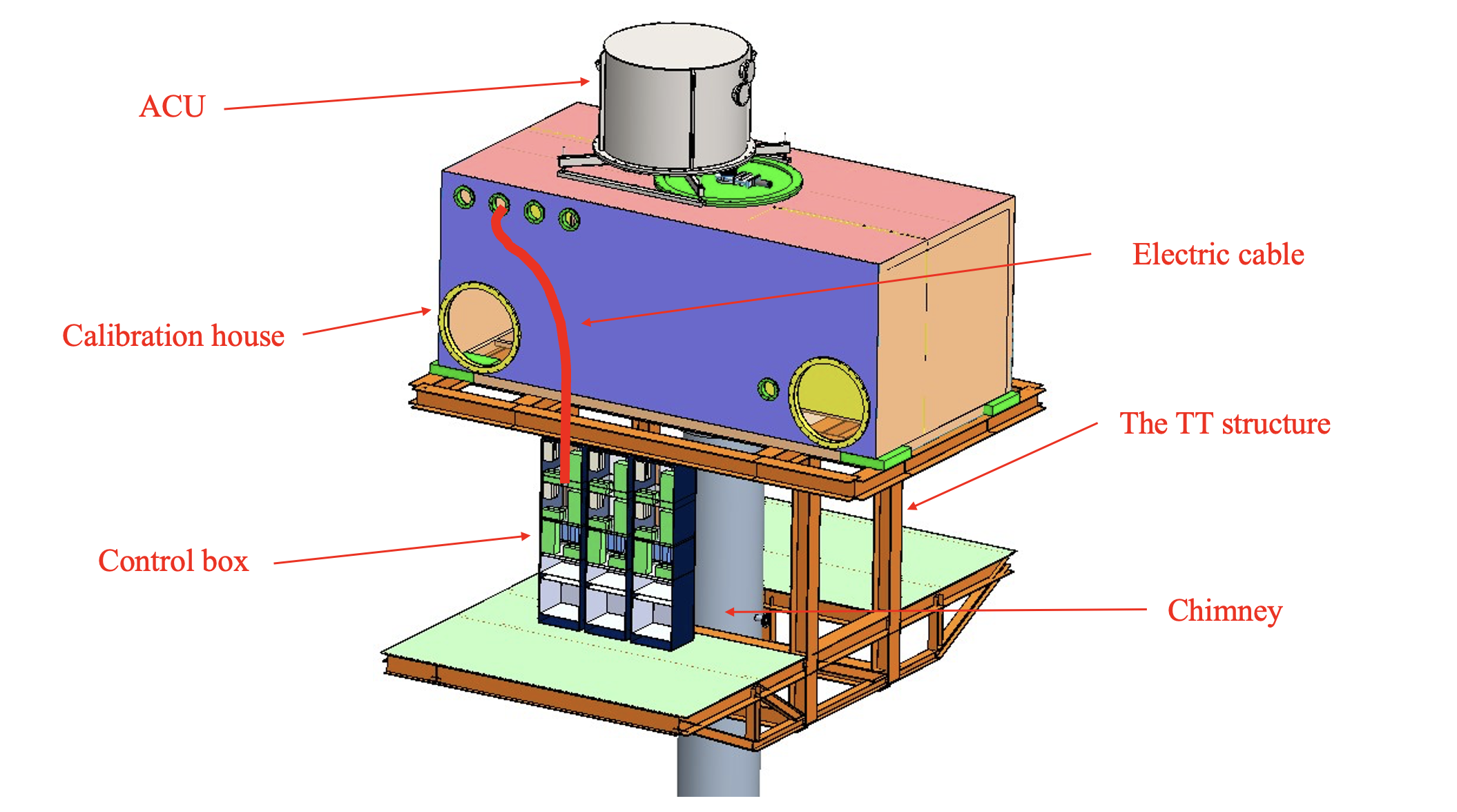}
  \caption{Illustration of the calibration house installed on the TT structure.}
  \label{calibration_house_outside2}
\end{figure}

\begin{figure}[!hbtp]
  \centering
  \includegraphics[width=0.8\textwidth]{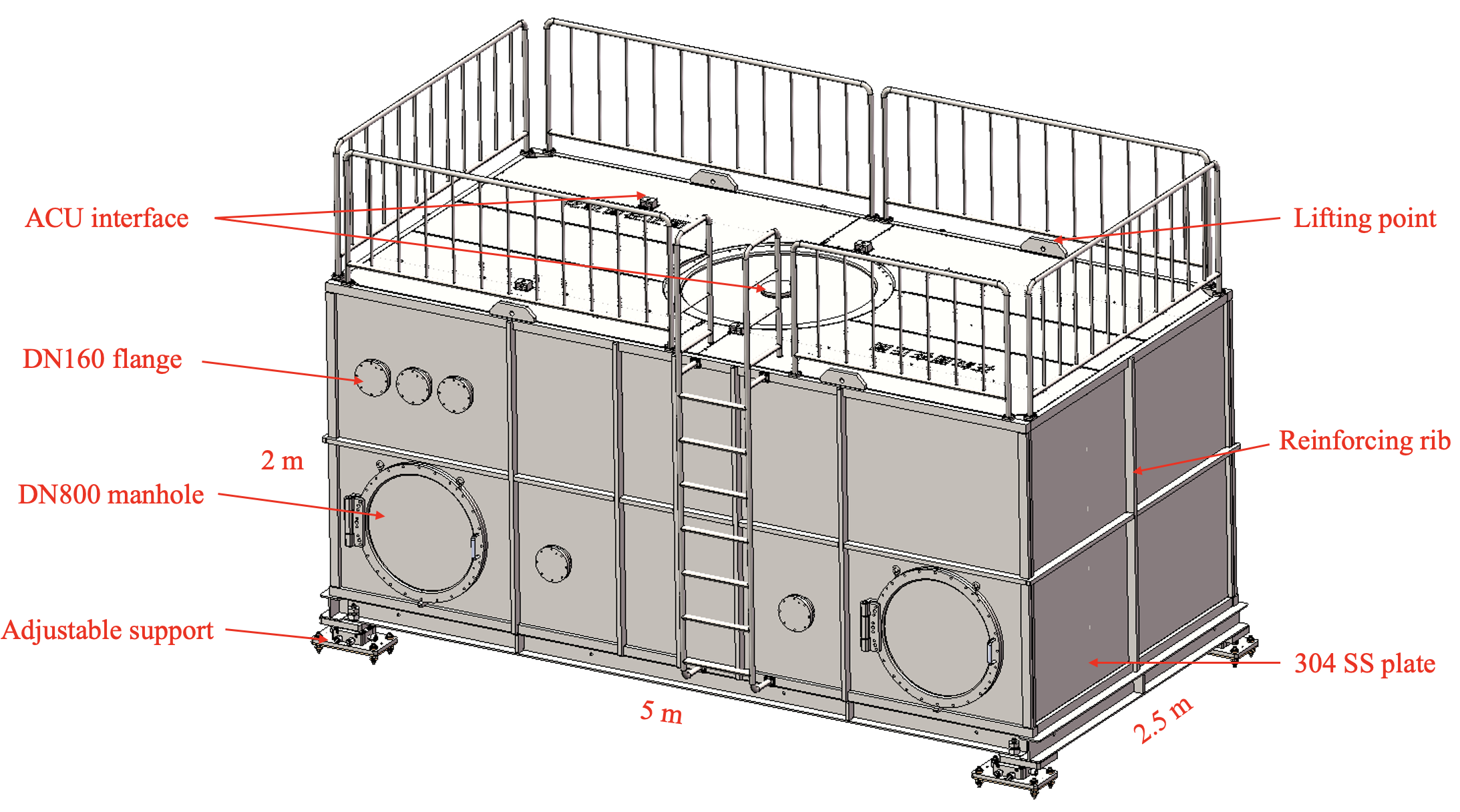}
  \caption{Illustration of the calibration house.}
  \label{calibration_house_outside1}
\end{figure}

The calibration house is an SS vacuum-sealed chamber with an internal dimension of 5~m~$\times$~2.5~m$\times$~2~m. The net weight of the calibration house is about 6~tons. The calibration house directly connects with the CD via a chimney. It is installed on the Top Track (TT) structure~\cite{Sandanayake:2024nbi}. Figure~\ref{calibration_house_outside2} shows the calibration house installed on the TT structure. As shown in Figure~\ref{calibration_house_outside1}, there is a 1250~mm inner diameter flange at the top of the calibration house that serves as the ACU interface and transportation hole for source storage, and a 820~mm inner diameter flange at the bottom that serves as the CD interface. Four DN150 flanges are installed at the four bottom corners of the calibration house used to flush nitrogen and connect signals through the Bayonet Nut Connector (BNC) feedthrough. Ten DN150 flanges and four DN800 flanges on the side wall of the calibration house serve as electric feedthrough flange and manhole, respectively. The electric feedthrough flange is designed to connect the CLS electronics inside the calibration house and its control box outside the calibration house. The manhole is used for transporting parts or devices into the calibration house and also serves as an entrance for personnel. 

\begin{figure}[!hbtp]
  \centering
  \includegraphics[width=0.8\textwidth]{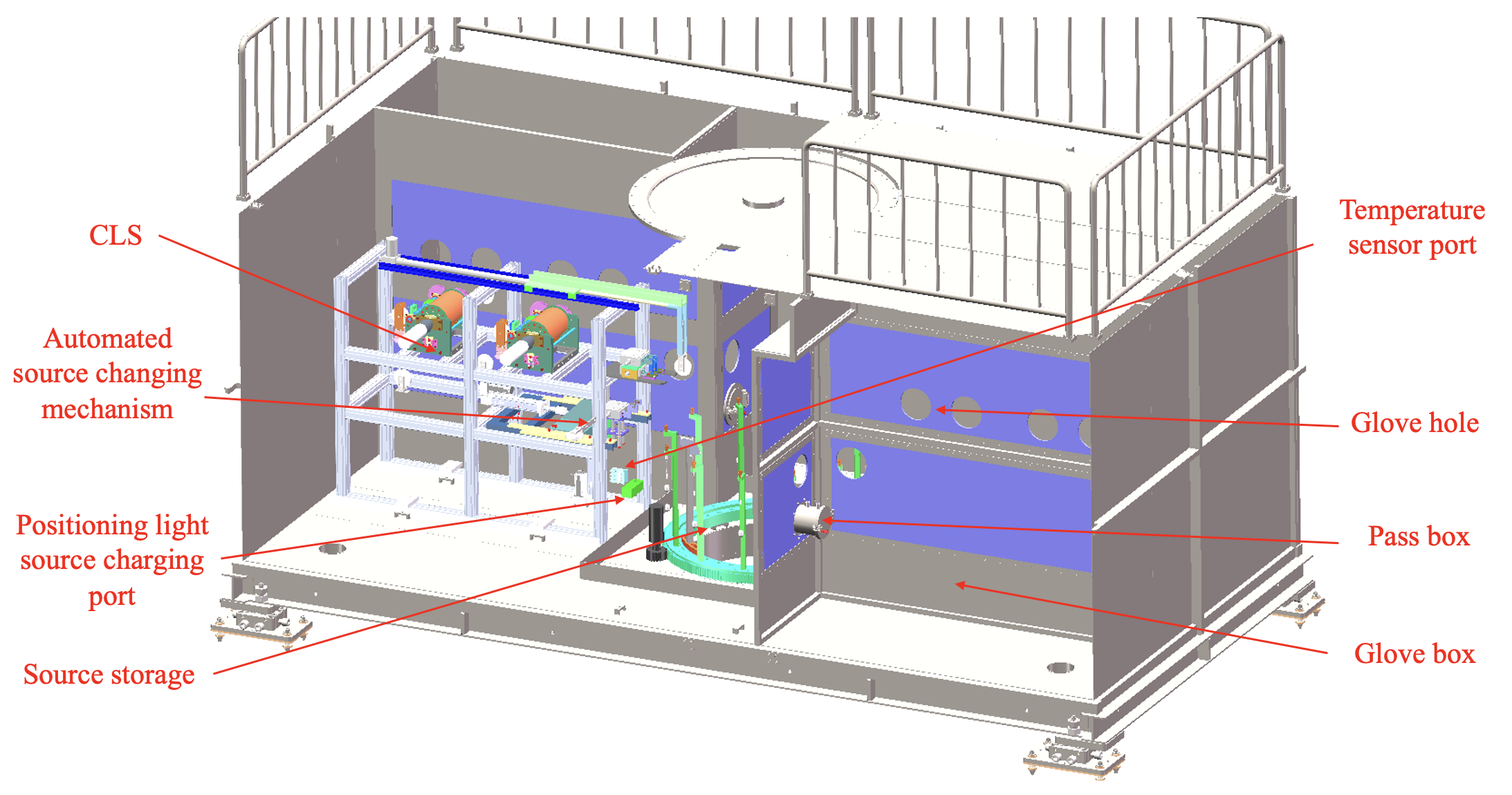}
  \caption{Illustration of the components installed inside the calibration house.}
  \label{calibration_house_inside}
\end{figure}

\begin{figure}[!hbtp]
  \centering
  \includegraphics[width=0.8\textwidth]{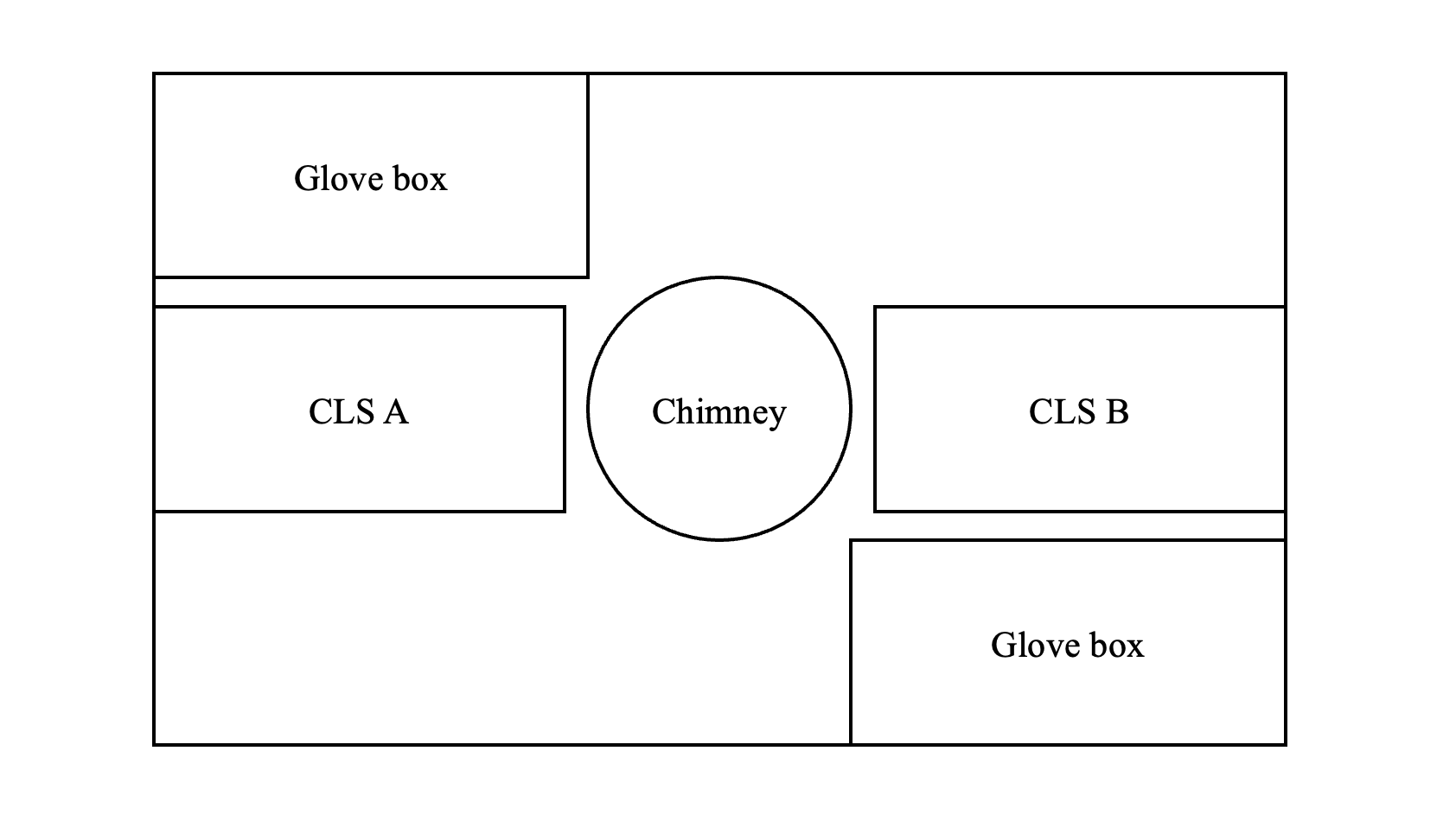}
  \caption{A schematic top view of the inner layout of the calibration house. The blank area next to the CLS is designated as the operation area for internal installation in the calibration house.}
 \label{calibration_house_top_view}
\end{figure}

Figure~\ref{calibration_house_inside} and Figure~\ref{calibration_house_top_view} show the components installed inside the calibration house and a schematic top view of the inner layout of the calibration house, respectively. It can be seen that inside the calibration house, there are two identical CLSs, two identical automated source changing mechanics, one source storage, and two identical glove boxes. Source storage, which is shared with two CLSs, is used to store radioactive sources, temperature sensor, positioning light source. Inside the calibration house fixed structures are reserved to install the CLS, source changing mechanic, and source storage.

\subsection{Design of the glove boxes}
\label{sec:design_glove_box}
The glove box is designed to fulfill two primary functions. First, it facilitates the transportation of radioactive sources or auxiliary devices through the pass box. Second, it allows the operator to deal with abnormal conditions. Detailed operations of the glove box are listed in Table~\ref{Glove_box_operation}. There are two identical glove boxes, each designated to cover a different area.

Each glove box is equipped with six gloves mounted onto a transparent acrylic panel, which also includes a pass box. The acrylic panel serves as both a viewport for direct observation of the CLS area during operations and a protective barrier. Under normal conditions, the glove box area is sealed with a DN800 manhole flange, which is opened only when access to the glove box is required for specific operations.

Table~\ref{Glove_box_operation} lists all the glove box operations and test results. It can be seen that the glove box can meet our requirement.

\begin{table}[!htp]
\centering
\caption{\label{Glove_box_operation} The glove box operation items and test results.}
\smallskip
\begin{tabular}{lll}
\hline
Test item                                                   & Motivation                                                                   & Result \\ \hline
Transport source and \\auxiliary device via pass box          & Share source or repair auxiliary device                                      & PASS   \\ 
Manually change source                                      & Automatic source changing mechanic malfunction & PASS   \\
Reset electric gripper                                      & Electric gripper gets stuck                                                  & PASS   \\
Reset the CLS SS cable                                          & The CLS SS cable jump out of the spool groove                                    & PASS   \\
Manually set CLS upper\\ pulley align with \\the CD center axis & The CLS upper linear motor malfunction                                           & PASS   \\
Change the USS emitter                                          & The USS emitter malfunction                                                  & PASS   \\
Manually operate temperature\\ sensor and light source        & Source change mechanic malfunction                                          & PASS   \\ \hline
\end{tabular}
\end{table}

To better understand Table~\ref{Glove_box_operation}, here we briefly introduce auxiliary device, CLS upper pulley, and USS emitter. 
Auxiliary device contains: temperature sensor for the CD temperature monitoring~\cite{Hui:2021dnh}; positioning light source, which serves as the CLS positioning system. Both temperature sensor and light source can be operated by source changing mechanic. The CLS upper pulley is mounted on a linear motor, which can align the pulley with the CD center axis during the CLS calibration. The USS emitter is a key component of the USS system, which is also the CLS positioning system.

\subsection{Distribution of electric cables and grounding}
\label{sec:electric_cable}
As two sets of CLSs and source changing mechanics are identical, this section introduces the electric cable distribution for one set. The electronic devices of the CLS, source changing mechanic, and other components are listed in Table~\ref{Electronic}. All electronics inside the calibration house must connect to their respective drivers or power supplies, which are located outside the calibration house. To maintain the air-tightness of the calibration house, feedthroughs (LEMO) are used for these connections.

\begin{figure}[!htbp]
\centering
\includegraphics[width=1.0\textwidth]{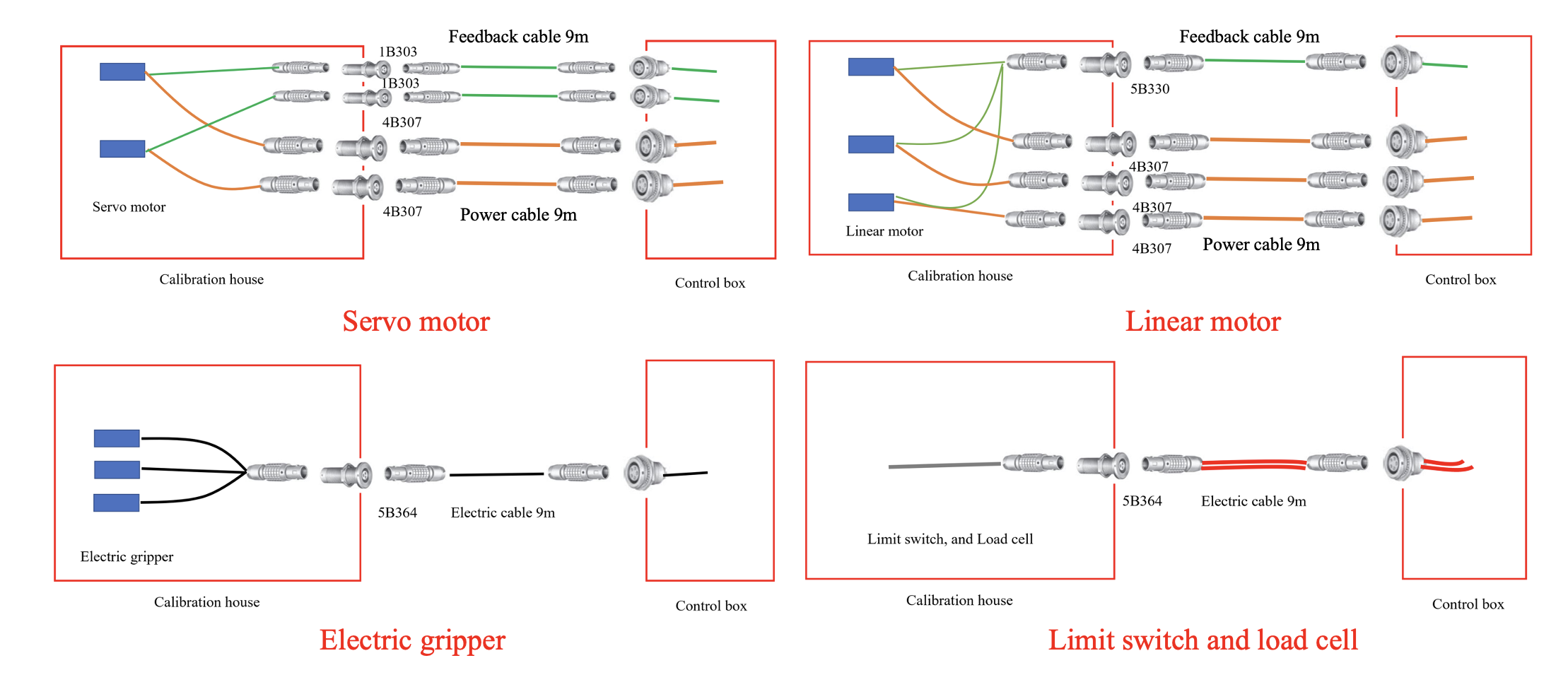}
\caption{Schematic diagram of the calibration house electric cable distribution. The yellow and green cables represent the servo motor power and feedback cables, respectively. The black cables represent the electric gripper cables, while the red cables represent the limit switch and load cell cables. The colors in the diagram match the actual cable colors to facilitate installation and feedthrough connection.}
\label{calibration_house_electric_cable_distribution}
\end{figure}

Cables are carefully distributed to minimize crosstalk and simplify installation. The schematic diagram of the calibration house’s electric cable distribution is shown in Figure~\ref{calibration_house_electric_cable_distribution}. Each motor (servo and linear motors) has two independent feedthroughs (distance larger than 50~mm): one for the power cable and another for the signal cable. The monitoring CCD camera shares the same BNC feedthrough as the ACU for signal transmission~\cite{Hui:2021dnh}.

All other electric cables, which transmit DC voltage, switching signals, and RS485 signals, are designed with strong anti-interference capabilities. These cables are bundled together and routed through a single feedthrough, streamlining the installation process and enhancing the overall tidiness of the setup.

\begin{table}[!htp]
\centering
\caption{\label{Electronic}The devices for each CLS, each source changing mechanic, and other associated components.}
\smallskip
\begin{tabular}{llll}
\hline
Device & Number & Function & Number of cables \\ \hline
Servo motor & 3 & Driven spool and source storage & 8*3 \\
Linear motor & 3 & Source change & 8*3 \\
Electric gripper & 3 & Source change & 24*3 \\
Load cell & 2 & Monitor tension of the CLS SS cable & 4*2 \\
Limit switch & 8 & Zeroing and position limitation & 8*2 \\ 
CCD cameras & 10 & Monitoring & 4*10 \\
Temperature sensor \\ readout port & 2 & For temperature sensor & 4*2 \\
Light source \\ readout port & 2 & For light source & 4*2 \\ 
Oxygen sensor & 1 & Monitor oxygen concentration & 4*2 \\\hline
\end{tabular}
\end{table}

\begin{figure}[!htbp]
\centering
\includegraphics[width=1.0\textwidth]{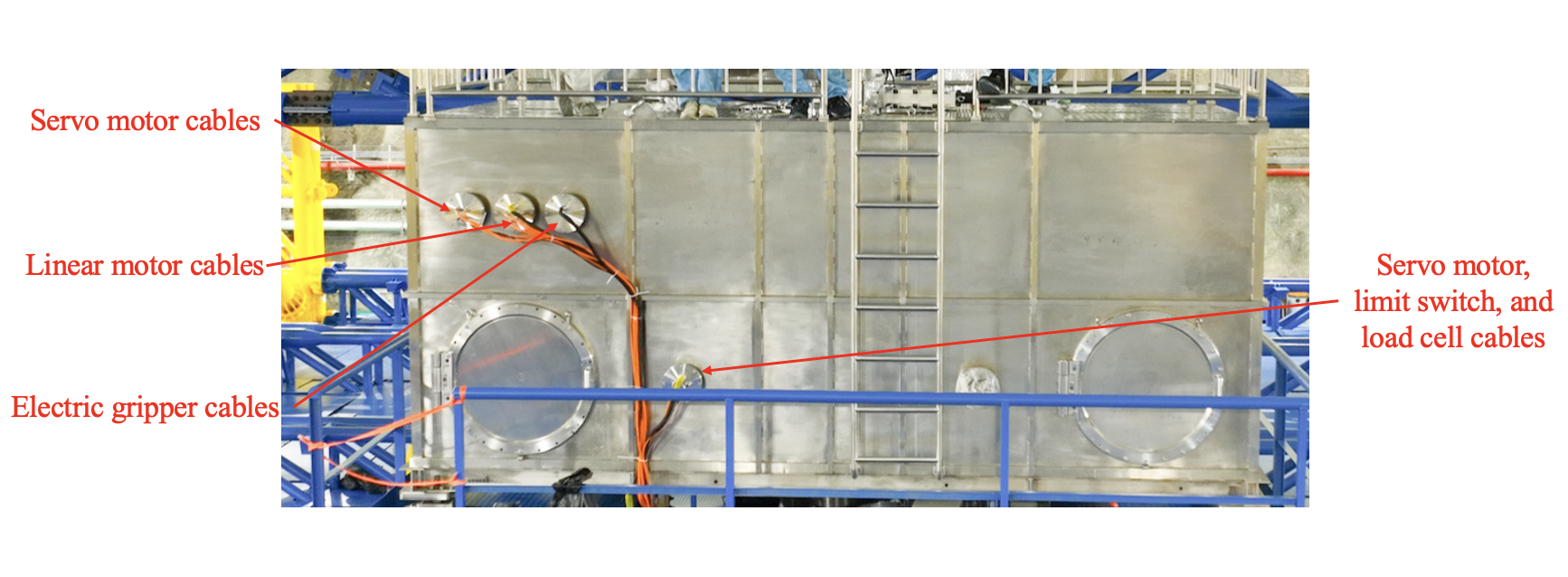}
\caption{Photo of distribution of the calibration house electric cables.}
\label{calibration_house_electric_cable_distribution_pic}
\end{figure}

Following the distribution strategy, Figure~\ref{calibration_house_electric_cable_distribution_pic} shows the calibration house electric cable distribution .

\begin{figure}[!htbp]
\centering
\includegraphics[width=0.8\textwidth]{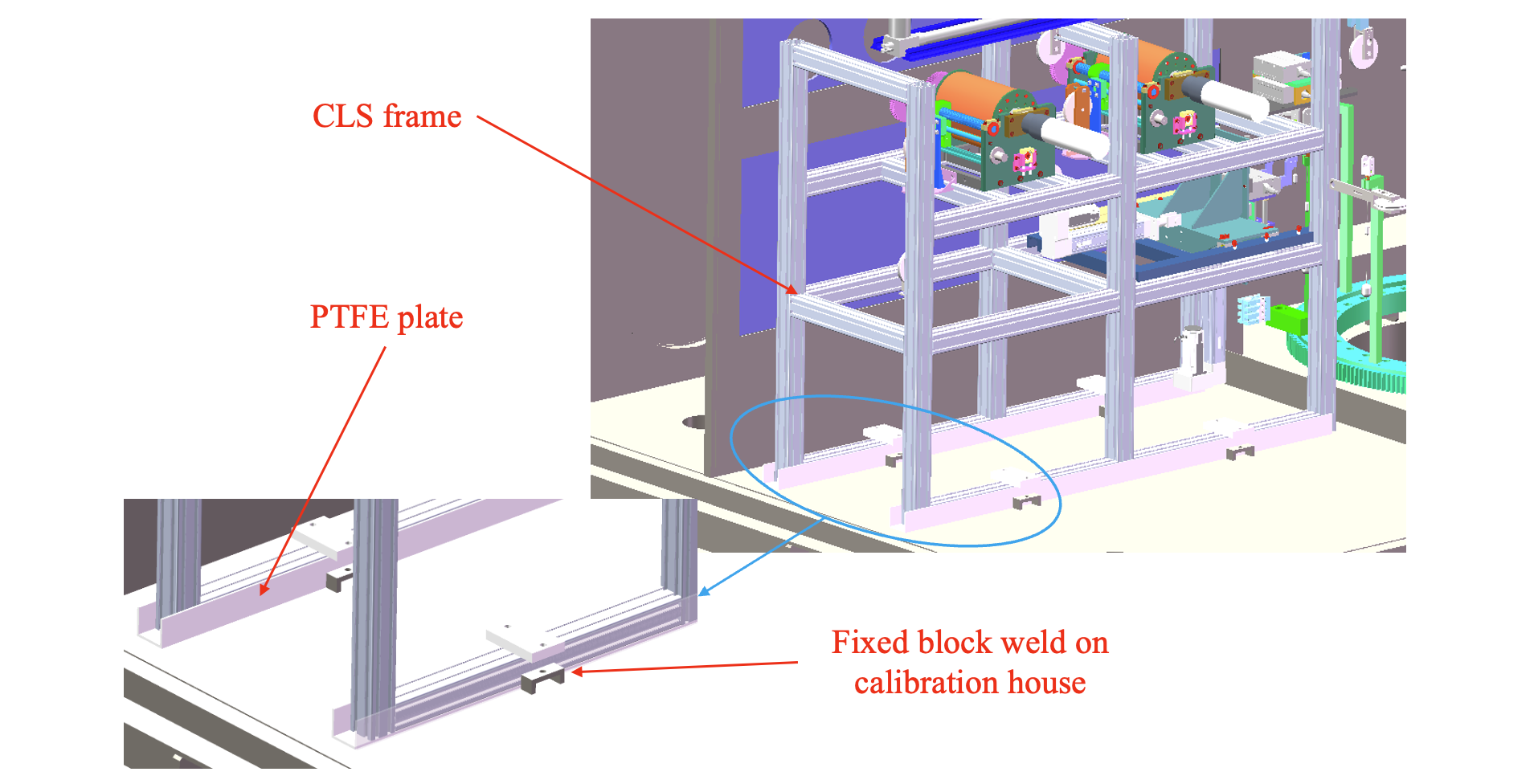}
\caption{An illustration of the insulation plan for the electronic devices of the CLS.}
\label{CLS_grounding}
\end{figure}

JUNO has two independent power and ground systems: the so called “clean” and “dirty”, considering noise level~\cite{grounding}. The PMT and its electronics system run on the clean power and clean ground, whereas most of the utilities run on the dirty power and safety ground. Since the CLS uses motors, by default, the CLS connected to the dirty power. On the other hand, the most CD structures are on the clean ground, therefore it is imperative that the electronics components inside the CLS be completely insulated from the calibration house. To achieve this, 5~mm thick Polytetrafluoroethylene (PTFE) plates, as illustrated in Figure~\ref{CLS_grounding}, are inserted between the CLS frame and calibration house. The source storage of the automated source changing mechanic is installed at the bottom of the calibration house using polyamide-imide bolts, which insulates electronic devices of source storage from the calibration house.

After assembling all the electric cables, the electronics inside the calibration house, the feedthroughs, and the control box, we conducted the CLS motion functionality test 100 times. No errors occurred during the tests, demonstrating that the electric cable distribution strategy meets our requirements.

\subsection{Nitrogen flush system}
\label{sec:design_oxygen_radon_control}
Since the calibration house is directly connected to the CD, the radon and oxygen concentrations in the calibration house will directly determine the radon and oxygen concentrations inside the CD. Radon and its decay products contribute to the radioactive background of the CD, while oxygen can affect the quality of the LS. Therefore, controlling the radon and oxygen concentrations in the calibration house is crucial. A nitrogen flush system is designed to remove radon and oxygen from the calibration house.

The main sources of oxygen in the calibration house are as follows: residual oxygen inside the calibration house after installation, air leaking from the outside, and oxygen entering from glove box leakage during operations. 

Similar to oxygen, radon in the calibration house originates from several sources: residual radon inside the calibration house after installation, air leaking from the outside, glove box leakage during operations, and radon emanating from materials inside the calibration house.

The nitrogen flush system comprises a High Purity Nitrogen (HPN) source, an inlet flush pipe, and an outlet flush pipe. Onsite, the HPN is supplied by the JUNO nitrogen system~\cite{Ling:2024gap}. During laboratory tests, pressurized HPN and boil-off liquid nitrogen, with an oxygen concentration of less than 5~ppm, are utilized as the HPN source. Both the inlet and outlet flush pipes are constructed using Polyurethane (PU) piping, featuring an inner diameter of 12~mm and an outer diameter of 16~mm. These pipes are connected to the calibration house via 1/2-inch SS piping welded onto the flushing flange.

\begin{figure}[!hbtp]
\centering
\includegraphics[width=0.8\textwidth]{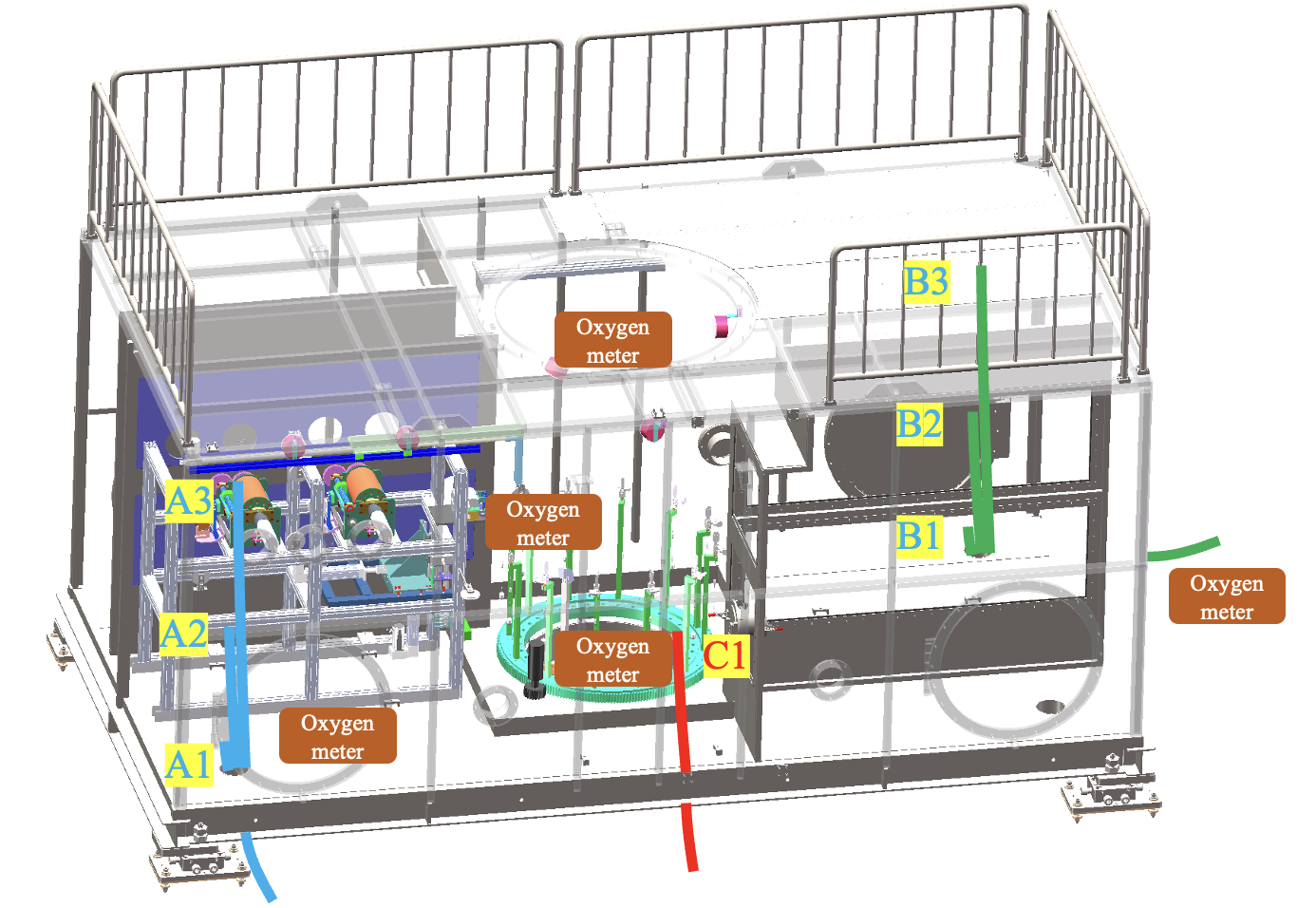}
\caption{Five different flush pipe layouts. Layout 1, input: A1, A2, A3; outlet: B1, B2, B3. Layout 2, input: A1; outlet: B3. Layout 3, input: A1; outlet: B1. Layout 4, input: A1, A2, A3; outlet: B1. Layout 5, input: C1; outlet: A1-A3 and B1-B3. Note that both the A1-A3 and outlet B1-B3 are connected to a single 1/2-inch pipe using a 4-way connector.}
\label{calibration_house_flush_pipe_layout}
\end{figure}

The positions of the nitrogen flush inlet and outlet pipes play a crucial role in determining the efficiency of oxygen and radon removal within the calibration house. Owing to the complex geometry inside the calibration house, determining the optimal layout for the flush pipes based solely on simulation is challenging. To address this, we have devised five different layouts based on past experience, as depicted in Figure~\ref{calibration_house_flush_pipe_layout}.

Layout 1 involves nitrogen flush from points A1, A2, and A3, with flush out from outlet points B1, B2, and B3. Layout 2 entails nitrogen flush from point A1, with flush out from point B3. Layout 3 comprises nitrogen flush from point A1, with flush out from point B1. Layout 4 involves nitrogen flush from points A1, A2, and A3, with flush out from point B1. Finally, layout 5 features nitrogen flush from point C1, with flush out from points A1-A3 and B1-B3. Note that both the A1-A3 and outlet B1-B3 are connected to a single 1/2-inch pipe using a 4-way connector. 

To determine the most effective flush pipe layout among these options, we utilized the oxygen concentration inside the calibration house as a figure of merit. To mitigate potential shortcuts or biases in the flush test, we employed five portable oxygen meters (ADKS-1-O2, range: 0-30$\%$VOL, accuracy: $\pm$~5$\%$ FS) to measure the oxygen concentration at various positions inside the calibration house. The positions of these oxygen meters are depicted in Figure~\ref{calibration_house_flush_pipe_layout}.

For each layout test, approximately 15~m$^3$ HPN is used and the pressure inside calibration house maintains to approximately 1600~Pa gauge pressure (measured by HALO-XQ-WG, 0-5000~Pa and MDM3051S series) during nitrogen flush. For layout 1-4, the flow rate is approximately 120~L/min, while for layout 5, the flow rate is approximately 350~L/min.

\begin{figure}[!htbp]
\centering
\includegraphics[width=0.9\textwidth]{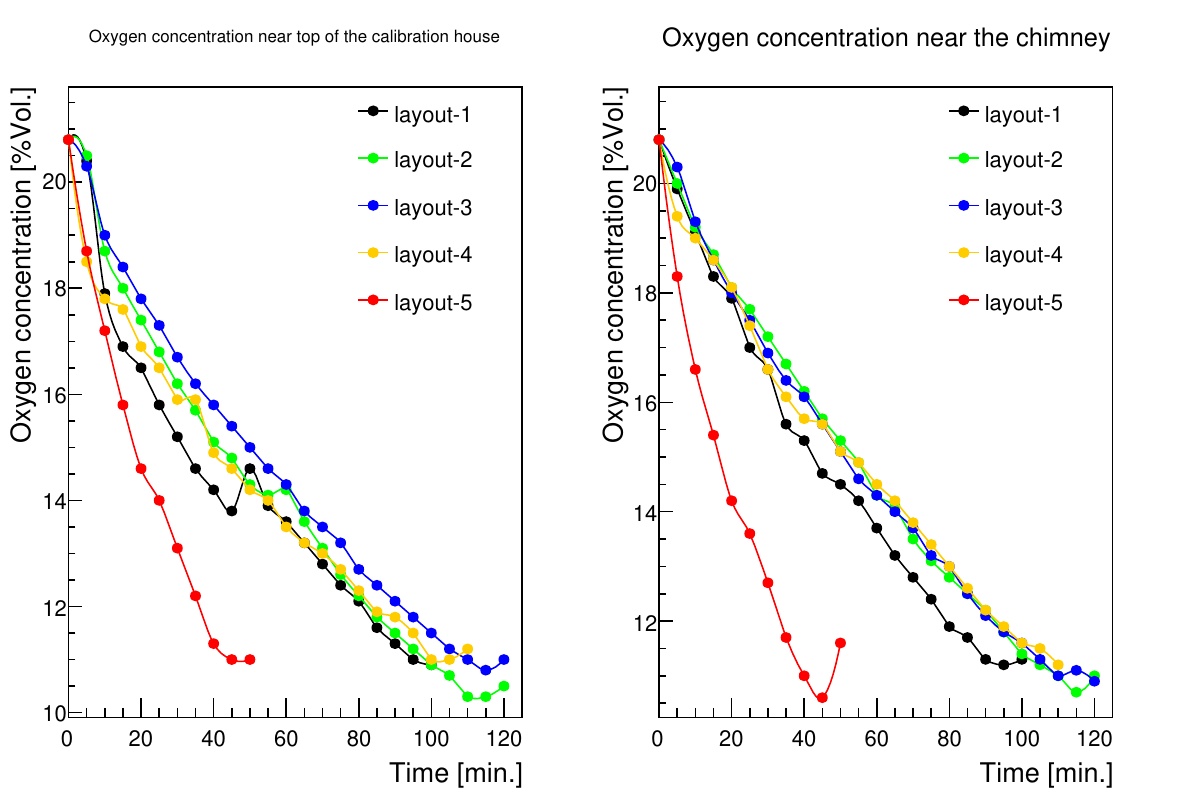}
\caption{Oxygen concentration at position near the chimney and near top of the calibration house with all the five flush pipe layouts. Black, green, blue, orange, and red colors indicate flush pipe layout 1, 2, 3, 4, and 5, respectively.}
\label{calibration_house_oxygen_concentration}
\end{figure}

Figure~\ref{calibration_house_oxygen_concentration} shows oxygen concentrations at the position near the chimney and near top of the calibration house with all the five flush pipe layouts. Black, green, blue, orange, and red colors indicate flush pipe layout 1, 2, 3, 4, and 5, respectively. It can be seen that the layout 5 has the best oxygen removal efficiency compare to the other layouts. For layout 5, it has two 1/2-inch outlet pipes rather than one; therefore, the flush time is half of other layouts. The oxygen concentration has increased at the end of the flush test because after running out of nitrogen, we keep to record oxygen concentration until the pressure inside the calibration house becomes lower than 100~Pa gauge pressure. During this time, residual oxygen in the hollow structure of some parts, such as the PTFE spool, inside the calibration house will diffuse out and increase the oxygen concentration. For layout 1, 2, and 4, it can be seen the little bumps in the middle of the test in Figure~\ref{calibration_house_oxygen_concentration}. This is because we stop flush and change the HPN bottles.

Since the geometry of the glove box is relatively simple, we determine the flush pipe layouts based on the basic principle, which is to maintain inlet away from outlet to avoid shortcut of gas flow.

\section{Background control of the calibration house}
\label{sec:test}

\subsection{Oxygen concentration control}
\label{sec:oxygen_control}
Oxygen concentration inside the calibration house is required to be less than 10~ppm. As discussed in Section~\ref{sec:design_oxygen_radon_control}, the sources of oxygen include residual oxygen, air leaking from the outside, and oxygen entering from glove box leakage during operations. 

For residual oxygen, nitrogen is flushed into the calibration house (flow rate: approximately 200~L/min, inside pressure: approximately 1600~Pa gauge pressure), and oxygen concentration is monitored near the chimney area. The oxygen concentration inside the calibration house can be reduced to $<$~5~ppm (measured by MOT500-IF-02 0-10000 ppm; with an accuracy better than 5~ppm around 10~ppm oxygen concentration), effectively keeping this source of oxygen under control.

For oxygen from air leaking from the outside of calibration house, leak rate of the calibration house is less than 4~$\times$~$10^{-5}$~mbar$\cdot$L/s. Therefore, this part of oxygen is negligible.

\begin{figure}[!htbp]
\centering
\includegraphics[width=0.8\textwidth]{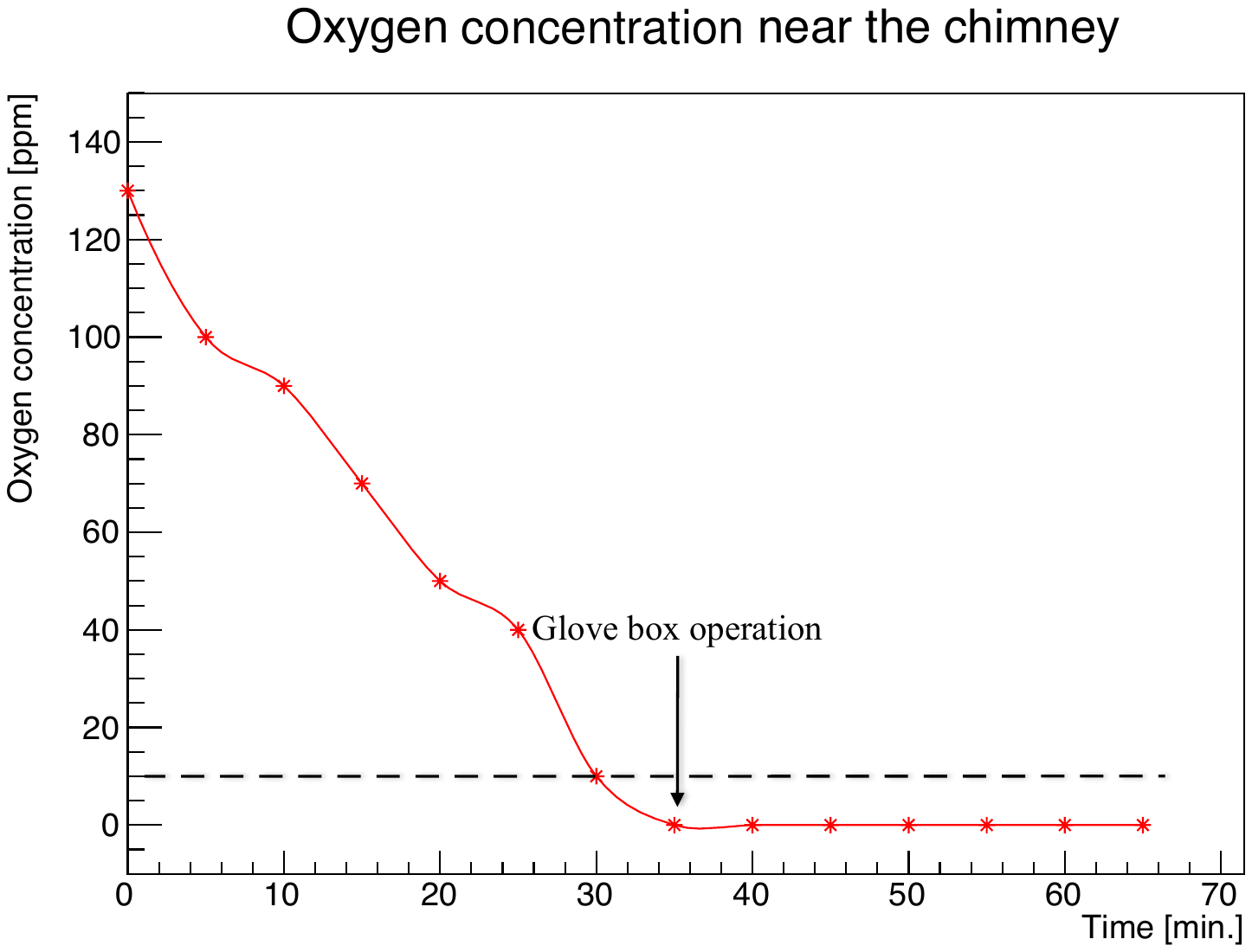}
\caption{Oxygen concentration at position near the chimney during glove box mock-up operation.}
\label{calibration_house_oxygen_leakage}
\end{figure}

Regarding oxygen leakage through the glove box during operations (when the manhole is opened), the leakage rate between glove box and the CLS area measured to be less than 1~mL/s level (the connections between the gloves and the acrylic panel, as well as the connection between the acrylic panel and the SS structure, could not ensure a vacuum-level seal). Oxygen leakage is controlled by, first, during glove box operations, higher pressure is maintained in the CLS area compared to the glove box area. Second, limiting the duration of each glove box operation (30 minutes is typical for a glove box operation). To evaluate the influence of glove box operation on oxygen concentration, we conducted a laboratory simulation while continuously monitoring the oxygen concentration within the CLS area. As shown in Figure~\ref{calibration_house_oxygen_leakage}, once the oxygen concentration inside the calibration house decreased to below 5~ppm, we initiated simulated glove box operations and continued nitrogen flushing for approximately 30 minutes. The results indicate that, within the sensitivity limits of the oxygen sensor, no detectable oxygen leakage into the CLS area occurred during glove box operation.

\subsection{Radon concentration control}
\label{sec:radon_control}
Radon concentration inside the calibration house must be controlled. The radon level in the LS must be less than 10$^{-9}$~Bq/L~\cite{JUNO:2021kxb}. Given that the total weight of the LS is 20~kton, the allowable radon concentration in the LS is calculated as follows:

$10^{-9}~\text{Bq/L}~\times~2.3~\times~10^{7}~\text{L} = 0.023~\text{Bq}$.

The radon decay product, $^{210}$Pb, has a long half-life (22.3 years) and its decay products contribute to the radioactive background. For solar neutrino studies, the $^{210}$Pb in the LS must be less than:

$10^{-24}~\text{g/g}~\times~20~\text{kton}~/~210~\text{g/mol}~\times~6.02~\times~10^{23}~\text{/mol}~\times~\ln2~/~22.3~\text{y} = 0.056~\text{Bq}$~\cite{JUNO:2021kxb}.

Therefore, considering both the radon and lead concentration requirements, the radon level inside the calibration house, which could diffuse into the CD via the chimney, is set to be less than 0.005~Bq (assume calibration house contribute approximately 20\% total budget), equivalent to 0.2~mBq/m$^3$.

As discussed in Section~\ref{sec:design_oxygen_radon_control}, radon in the calibration house originates from several sources: residual radon remaining after installation, calibration house leakage, glove box leakage during operations, and radon emanation from materials inside the calibration house.

For the first three sources, the same control measures used for oxygen are applied, and any residual radon will naturally decay over time. Additionally, radon-free air is supplied to the glove box area during glove box operations to ensure safe breathing conditions for the operators. Therefore, radon leakage from the glove box during operations is negligible. To minimize radon emanation from materials inside the calibration house, low-background materials are selected, the inner surfaces of the calibration house are electrolytic polished, and all equipment inside the calibration house is cleaned.

Radon emanating from materials inside the calibration house enters the CD through the chimney. However, due to the short half-life of $^{222}$Rn (3.82~days), not all radon atoms have chance to enter the chimney before decaying. Therefore, we cannot directly apply the radon concentration requirement for the detector as the same requirement for the calibration house; instead, there is a conversion relationship between the two. We establish a relatively simple diffusion model to estimate the radon concentration within the calibration house in relation to the radon actually entering the detector. The following two subsections will introduce two different diffusion models.

\subsubsection{Radon diffusion model 1}
\label{sec:radon_model1}
Assuming the amount of radon diffuse into the CD is proportional to the surface area of the LS in the chimney and the calibration house inner surface. This means that radon moves freely within the calibration house, and when it collides with the inner surface of the house before decaying, it gets bounced back. However, if it collides with the LS inside the chimney, it will be absorbed and enter the CD. The area of the calibration house inner surface is approximately 55~m$^2$. And the area of the LS in the chimney is approximately 0.5~m$^2$. Therefore, about 1\% radon inside the calibration house has chance diffusing into the CD.

\subsubsection{Radon diffusion model 2}
\label{sec:radon_model2}
The second radon diffusion model considers the radon concentration at the bottom of the chimney (before entering the acrylic sphere of the CD) during 20 years. Considering that no convection occurs in the upper 4-meter-deep LS within the chimney during stable data acquisition, since this section is outside the water pool and experiences minimal heat exchange, radon is not transported from the calibration house to the bottom of the chimney by convection. Therefore, we assume that radon transport follows the diffusion model described in reference~\cite{d}. Although radon has a very short lifetime, the long lifetime of its decay products can still impact the background of the detector. Assuming that radon and its decay products share the same diffusion coefficient. The ratio of radon and its decay products concentration at the 4-meter-deep LS of the chimney to the radon concentration inside the calibration house $R$ can be expressed as: 

\begin{equation}
R=erfc(\frac{h}{\sqrt{4a\tau}}), \label{eq:3}
\end{equation}

where, the $erfc$ is complementary error function, the $h=4$~m is the width of the LS, the $a$ is the diffusion coefficient which is determined to be 1.49$\times 10^{-9}$~m$^2$/s based on the study in reference~\cite{Xu:2023smq}, and the $\tau$ is the diffusion time, which is considered as 20 years. Therefore, $R \approx 0.35\%$, indicating that, without convection, radon and its decay products can only diffuse through the 4-meter LS to reach about 0.35\% of the surface concentration at the bottom over 20 years.

\subsubsection{Measurement of radon concentration inside calibration house}
\label{sec:radon_concentration}
The radon concentration in the calibration house is measured using a customized radon detector~\cite{PandaX-4T:2021lbm}. By applying the coefficient that accounts for radon entering the detector from the calibration house, we can determine whether the radon concentration meets the required standards. Since the primary source of radon in the calibration house is the materials within the house, radon levels can be reduced by using materials or equipment with lower radon emissions.

\begin{figure}[!htbp]
\centering
\includegraphics[width=0.8\textwidth]{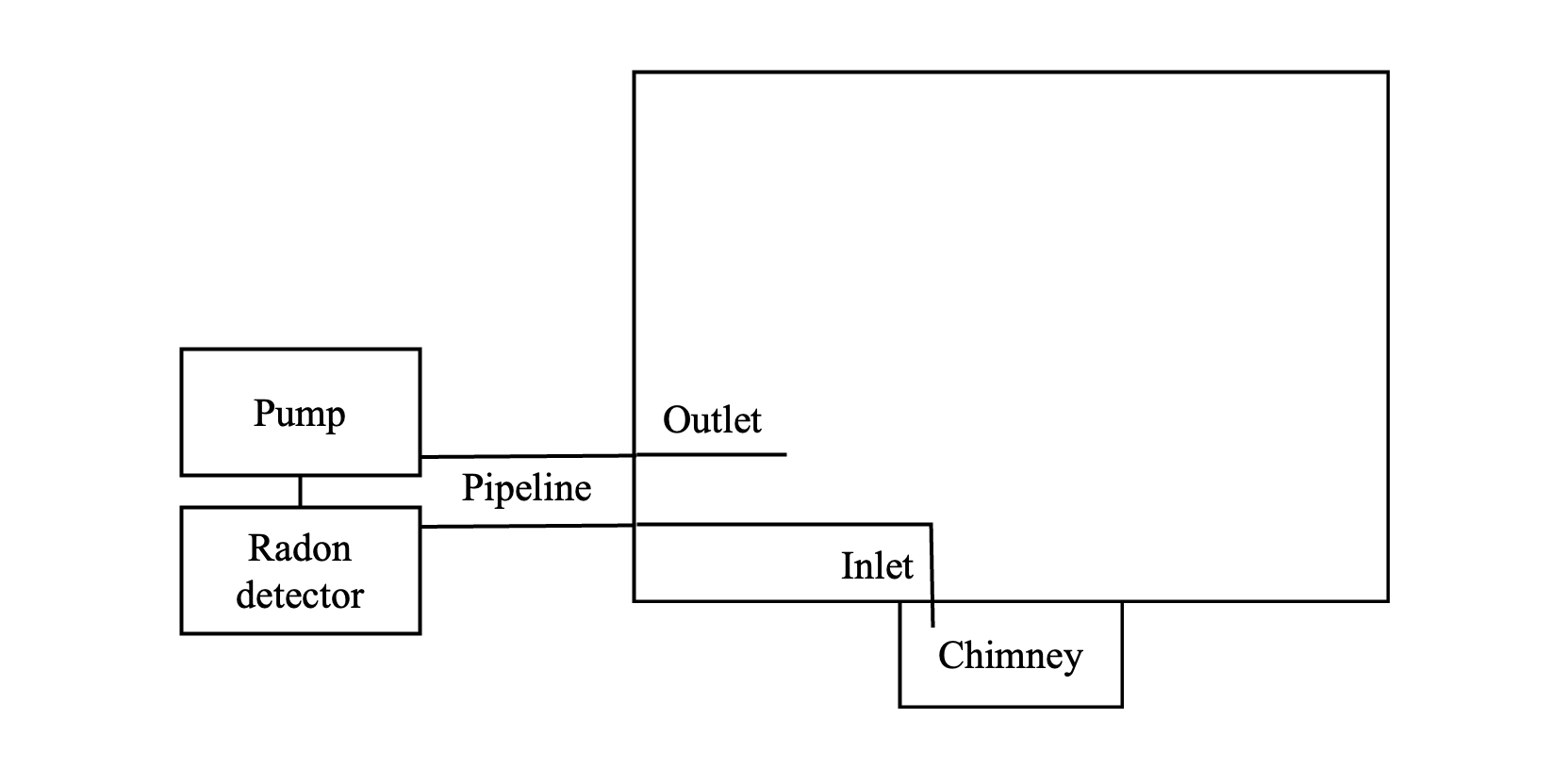}
\caption{Schematic diagram of the calibration house radon concentration measurement setup.}
\label{radon_measurement_setup}
\end{figure}

To identify the main sources of radon emissions from different components (CLS, calibration house, and dust), multiple measurements were conducted. Figure~\ref{radon_measurement_setup} shows the schematic diagram of the calibration house radon concentration measurement setup. It mainly consists of a customized radon detector, a pump, and pipelines. Comparing these measurements allows us to pinpoint the primary sources of radon and decide whether it is necessary to replace materials or equipment with lower radon emissions.

In the first measurement, the uncleaned calibration house and CLS were sealed, and the radon concentration in the calibration house stabilized at 50~$\pm$~10~mBq/m$^3$. The total uncertainties reflect the propagation of statistical errors from both signal and background contributions, combined with a 5\% detector-related systematic uncertainty~\cite{ra}.

In the second measurement, after cleaning both the calibration house and CLS, the radon concentration decreased to [11, 46]~mBq/m$^3$ at 90\% Confidence Level (C.L.) (28~$\pm$~11~mBq/m$^3$), indicating that dust contributed 22~$\pm$~15~mBq/m$^3$ of radon~\cite{PhysRevD.57.3873}.

In the third measurement, the CLS was removed, and the radon concentration in the calibration house dropped to $<$~11~mBq/m$^3$ at 90\% C.L. (3~$\pm$~5~mBq/m$^3$), suggesting that the CLS contributed approximately 25~$\pm$~12~mBq/m$^3$ of radon, while the calibration house itself contributed less than 11~mBq/m$^3$.

In the fourth measurement, nitrogen flush was introduced based on the second measurement setup, resulting in a radon concentration of [2, 15]~mBq/m$^3$ at 90\% C.L. (8~$\pm$~4~mBq/m$^3$). This indicates that nitrogen flush is able to suppress approximately 3 times the radon concentration in the calibration house. Nitrogen flush has different removal efficiencies for oxygen and radon. Oxygen concentration decreases from 20\% to 1~ppm, while radon concentration drops from 28~$\pm$~11~mBq/m$^3$ to 8~$\pm$~4~mBq/m$^3$. A possible reason is that the oxygen concentration in the calibration house is more than 16 orders of magnitude higher than the radon concentration.

It should be noted that changing the setups require opening the calibration house, which introduces radon from the air. Therefore, before the measurements, we flush the calibration house with nitrogen to reduce the oxygen concentration to below 5~ppm. After that, we wait for more than 10 days to allow the radon inside the house to reach equilibrium before starting radon measurements.

The conditions of the radon measurement and results are summarized in Table~\ref{radon}.

\begin{table}[!hbtp]
\centering
\caption{\label{radon} Summarized conditions of the radon measurement and results.}
\smallskip
\begin{tabular}{ccc}
\hline
Setup & Radon concentration (mBq/m$^3$) \\ \hline
Calibration house + w/ CLS + w/o cleaning + w/o nitrogen flush & 50~$\pm$~10 \\
Calibration house + w/ CLS + w/ cleaning + w/o nitrogen flush & [11, 46] at 90\% C.L.\\
Calibration house + w/o CLS + w/ cleaning + w/o nitrogen flush & $<$~11 at 90\% C.L.\\
Calibration house + w/ CLS + w/ cleaning + w/ nitrogen flush & [2, 15] at 90\% C.L.\\ \hline
\end{tabular}
\end{table}

Considering the coefficient that accounts for radon entering the detector from the calibration house in Section~\ref{sec:radon_model1} and Section~\ref{sec:radon_model2} and the radon concentration inside the calibration house (8~$\pm$~4~mBq/m$^3$), it can be seen that the radon concentration inside the calibration house is conservatively expected to be 0.08~$\pm$~0.04~mBq/m$^3$, which is smaller than our requirement (0.2~mBq/m$^3$).

\subsection{Onsite underground leak check}
\label{sec:leak_check}
The design leakage rate of the calibration house is 4~$\times$~$10^{-5}$~mbar$\cdot$L/s. Considering the large size of the calibration house, thin (2~mm) SS plates were used in the design to reduce the overall weight of the calibration house, so it is not possible to directly vacuum and leak check the entire calibration house due to its weak strength. In the laboratory, we conduct helium sniffing leakage check, and the leakage check result is less than 4~$\times$~$10^{-6}$~mbar$\cdot$L/s. But, leakage check is still required after the calibration house transportation from our laboratory to onsite underground, considering that all flanges need to be reinstalled. 

Unlike in the laboratory, the helium sniffing method is not suitable for onsite underground use due to the potential damage to the PMTs. Instead, we employ a different leakage check strategy. For single O-ring feedthroughs, we utilize the SF$_6$ sniffing leak detector (Kstone LF-301-H, with a calibrated sensitivity better than 1~$\times$~$10^{-6}$~mbar$\cdot$L/s). In the case of flanges, since they all feature double O-ring seals, we can fill the space between the O-rings with nitrogen gas and use a pressure gauge to measure the pressure drop. The leak rate can then be calculated using $dp/dt$, where $dp$ is the pressure drop within a given time interval $dt$. The leak check sensitivity of this method is better than 1~$\times$~$10^{-6}$~mbar$\cdot$L/s. A total of 21 flanges in the calibration house are tested using this method, with a leak rate requirement of less than 1~$\times$~$10^{-6}$~mbar$\cdot$L/s for each flange. Therefore, the total leakage from all flanges is less than 2.1~$\times$~$10^{-5}$~mbar$\cdot$L/s. Taking into account the leak test results of the single O-ring feedthroughs, which have a leakage rate below 0.8~$\times$~$10^{-5}$~mbar$\cdot$L/s, the total leakage rate of the calibration house is better than 2.9~$\times$~$10^{-5}$~mbar$\cdot$L/s.

This leak check strategy has been performed after calibration house installation and total leak rate is less than 2.9~$\times$~$10^{-5}$~mbar$\cdot$L/s, which meets our requirements.


\section{Summary}
\label{sec:conclusion}
This paper introduces the design and related testing of the calibration house, one of the auxiliary systems of the JUNO calibration system. The design section primarily covers mechanical design, electric cable distribution design, and the nitrogen flush system design. The testing section focuses on controlling the oxygen and radon concentrations inside the calibration house, glove box testing, and onsite leak check. Notably, the oxygen concentration in the calibration house can be controlled to be less than 10~ppm, and the radon concentration can be maintained at less than 15~mBq/m$^3$. Onsite installation and testing results demonstrate that the calibration house meets the requirements of the calibration system.

\section*{Acknowledgments}
The authors sincerely thank all JUNO collaborators for their valuable comments, as well as the CD and FOC (Filling, Overflowing
and Circulation System) groups for their support and insightful discussions during the installation and commissioning of the calibration house. Additionally, the authors would like to express their gratitude to the members of Shanghai Jiao Tong University.

This work is supported by the Strategic Priority Research Program of the Chinese Academy of Sciences (Grant Number: XDA10010800), the CAS Center for Excellence in Particle Physics (CCEPP), the Office of Science and Technology, Shanghai Municipal Government (Grant Number: 16DZ2260200), the Key Laboratory for Particle Physics, Astrophysics and Cosmology, Ministry of Education, the National Key Research and Development Program of China (Grant number: 2023YFA1606104), and the National Science Foundation of China (Grant number: 12250410235, 12222505). Y. M. and J. H. thank the sponsorship from the Yangyang Development Fund.

\bibliographystyle{JHEP} 
\bibliography{jinst-latex-sample}

\end{document}